\documentstyle[aps,prb,epsfig]{revtex}
\begin{document}
\renewcommand{\thefigure}{\arabic{figure}}
\baselineskip=0.5cm
\title{Analytic theory of correlation energy and spin polarization
in the $2D$ electron gas}
\author{R. Asgari,$^{a,b}$ B. Davoudi,$^{a,b}$ and M. P. Tosi$^a$~\footnote{Corresponding author: tosim@sns.it}}
\address{$^a$NEST-INFM and Classe di Scienze, Scuola Normale Superiore, I-56126 Pisa, Italy\\
$^a$Institute for Studies in Theoretical Physics and Mathematics, Tehran
19395-5531, Iran\\}
\maketitle
\vspace{0.2 cm}

{\bf Abstract}

We present an analytic theory of the pair distribution function and the ground-state energy 
in a two-dimensional ($2D$) electron gas with an arbitrary degree of spin polarization. Our approach involves the solution of a zero-energy scattering Schr\"odinger equation with an effective potential which includes a Fermi term from exchange and kinetic energy and a Bose-like term from Jastrow-Feenberg correlations. The form of the latter is assessed from an analysis of data 
on a $2D$ gas of charged bosons. We obtain excellent agreement with data from quantum Monte Carlo studies of the $2D$ electron gas. In particular, our results for the correlation energy show a quantum phase transition occurring at coupling strength $r_s\approx 24$ from the paramagnetic 
to the fully spin-polarized fluid.

\noindent PACS numbers: 05.30.Fk;  71.10.Ca

\noindent {\it Key Words:} D. Electron-electron interactions
\newpage

 Developments in techniques of molecular beam epitaxy and chemical vapor deposition 
have allowed since the mid 1970's the fabrication of semiconductor structures in which the 
carriers can form a low-density fluid moving in low dimensionality~[1]. Many of the electron-electron interaction effects become increasingly important as carrier density and dimensionality are reduced and the homogeneous electron gas - an assembly of fermions interacting by the $e^2/r$ 
law and moving in a uniform neutralizing background - provides a primitive model for their study~[2]. The model is characterized at zero temperature and magnetic field by the carrier density or equivalently by the coupling-strength parameter $r_s$, which measures the mean interparticle distance in units of the Bohr radius $a_B$.

A crucial role in the theory is played by the particle pair distribution function $g(r)$, which determines the ground-state energy of the assembly and hence its quantum phase diagram. This function has been the object of several numerical studies by quantum Monte Carlo (QMC) 
methods, starting from the early work of Ceperley and Alder~[3] on the three-dimensional ($3D$)electron gas. In the two-dimensional ($2D$) case these studies have predicted the critical density 
for Wigner crystallization at $r_s\approx 35$~[4,5] and a weakly first-order transition to a fully spin-polarized fluid state at $r_s \approx 25$~[6,7].

In a previous work we have presented an analytic theory of $g(r)$ and other ground-state properties of the $3D$ electron gas~[8]. The theory was based on a Fermi hypernetted chain approximation (hereafter indicated by the FHNC/0 acronym) and yields quantitative agreement 
with QMC data up to at least $r_s = 20$. The object of the present work is to extend the theory of $g(r)$ and the correlation energy to the $2D$ electron gas. As we shall see, much higher sophistication is needed to attain a quantitatively useful theory in lowered dimensionality. We 
shall have to dwell on terms beyond the FHNC/0, which come from low-order elementary 
diagrams and from three-body Jastrow-Feenberg correlations. These effects have been studied theoretically in boson fluids~[9,10] and it is natural within our theoretical scheme to draw on the $2D$ gas of charged bosons in dealing with the $2D$ gas of charged fermions. An important role will be played below by the newly available QMC data on the $2D$ charged-boson gas~[11].

	Our starting point is the energy functional $E[n(\bf r)]$ for an inhomogeneous density profile corresponding to the distribution of particles  $n(\bf r)$ around an average particle. That is, $n(\bf r)$$= ng(r)$ 
, where $n =(\pi r_s^2 a_B^2)^{-1}$ is the average particle density, and the corrsponding "external" potential is $v(r) = e^2/r$. As usual, the energy functional is the sum of an ideal kinetic energy term and of three potential energy terms given by the interaction with the "external" potential and by 
the Hartree and exchange-correlation contributions. As in Ref. [8] we split the ideal kinetic energy functional into the sum of the von Weizs\"{a}cker - Herring term ($\hbar^2/8 m) \int d{\bf r}\left| \nabla n(\bf r) \right|^2/ n(\bf r) $ and a residue $T_{\theta}[n(\bf r)]$ , and minimize the energy by means of the Hohenberg-Kohn variational 
principle. The result is a differential equation for $\sqrt{g(r)}$ ,
  
\begin{equation}
[-\frac{\hbar^2}{m}\nabla^2+v(r)+W_F(r)+W_B(r)]\sqrt{g(r)}=0~.
\end{equation}
This equation is formally exact and describes a zero-energy two-body scattering problem in the many-body fluid, in accord with Overhauser's interpretation of $g(r)$~[12].

In writing the "scattering potential" in the form used in Eq.~(1) we are implicitly thinking of the many-body wave function as the product of a Slater determinant times a Jastrow-Feenberg correlation factor. The Fermi potential $W_F(r)$ derives from the functional derivative of $T_{\theta}[n(\bf r)]$ and from exchange, and would vanish in a Bose fluid. The Bose-like potential $W_B(r)$ contains the effects of correlations (including the long-range Hartree term) and by itself would determine 
$g(r)$ in a Bose fluid. We turn to the latter in order to assess the form of $W_B(r)$~.

{\it $2D$ fluid of charged bosons}. The FHNC/0 and its HNC/0 equivalent for a Bose fluid take 

\begin{equation}
W_B^{HNC}(k)=-\frac{\hbar^2k^2}{4m}\left[\frac{S(k)-1}{S(k)}\right]^2[2S(k)+1]
\end{equation}
where $W_B(k)$ is the Fourier transform of $W_B(r)$, defined according to the general expression $FT[F(r)]= n\int d{\bf r} F(r) \exp{(i{\bf k}\cdot {\bf r})}$. In Eq.~(2) $S(k)$ is the structure factor of the fluid, defined so that $S(k)=1+ FT[g(r)-1]$. Using Eq.~(2) in Eq.~(1) with $W_F(r) = 0$ and self-consistently solving Eq.~(1) for $g(r)$, we find quantitative agreement with the QMC data of Ref.~[11] at $r_s = 1$ and discrepancies already emerging at $r_s = 5$~.

Improvements on Eq.~(2) can be sought in two directions for a Bose fluid ~[10,13]. The HNC may be transcended by the inclusion of low-order elementary diagrams, and three-body Jastrow-Feenberg correlations may be included. The contribution from the first elementary diagrams to the effective Bose potential is
\begin{equation}
W_B^{E_4}(k)=-\frac{\hbar^2}{4mn}\left\{\frac{k^2}{(2\pi)^2}\varepsilon_4(k)+\int\frac{d{\bf q}}{(2\pi)^2}q^2[S(q)-1]\frac{\delta\varepsilon_4(q)}{\delta S(k)}\right\},
\end{equation}
where $\varepsilon_4(k)$ is given by a fourfold integral (in $2D$) over momentum space. The contribution of 
three-body correlations is given by a two-fold integral,

\begin{eqnarray}
W_B^{(3)}(k)=\frac{1}{4n(2\pi)^2}\int d{\bf q}S(p)S(q)
u_3({\bf q},{\bf p},{\bf k})
\left\{v({\bf q},{\bf p},{\bf k})
+\left[\varepsilon(p)+\varepsilon(q)\right]u_3(\bf{q},\bf{p},\bf{k})\right\}
\end{eqnarray}
where ${\bf p}=-({\bf q}+{\bf k})$, $\varepsilon(k)=\hbar^2k^2/[2mS(k)]$ and, with the definition $X(k)=1-S^{-1}(k)$, we have $v({\bf q},{\bf p},{\bf k})=(\hbar^2/m) \left[{\bf k}\cdot{\bf p}X(p)+{\bf k}\cdot{\bf q}X(q)+{\bf p}\cdot{\bf q}X(p)X(q)\right]$ and

\begin{equation}
u_3({\bf q},{\bf p},{\bf k})=-\frac{(\hbar^2/2m)\left[{\bf k}\cdot{\bf p}X(p)X(k)+{\bf p}\cdot{\bf q}X(p)X(q)+{\bf k}\cdot{\bf q}X(q)X(k) \right]}{\varepsilon(k)+\varepsilon(p)+\varepsilon(q)}
\end{equation}
The papers of Smith {\it et al}.~[13] and of Apaja {\it et al}.~[10] should be consulted for detailed derivations of these equations, in which $u_3({\bf q},{\bf p},{\bf k})$ is the irreducible three-body vertex.

The effective scattering potential as self-consistently determined in these approximations for the $2D$ fluid of charged boson at $r_s = 20$ is shown in Fig.~1. In all cases the structure in $g(r)$ 
which is present in the QMC data of Ref.~[11] is underestimated at such strong coupling. We have therefore tried to take into account the higher-order terms that are missed in these approaches at strong coupling by assuming that they lead to corrections in the scattering potential that have roughly the same 
shape as the low-order terms reported above. In particular, we have found that the choice
\begin{equation}
W_B^{(\alpha)}(r)=W_B^{HNC}(r)+\alpha(r_s) W_B^{(3)}(r)
\end{equation}
leads to a satisfactory account of the QMC data on the $2D$ Bose gas. Here the parameter $\alpha$ is determined by fitting the QMC data on the ground-state energy~[11] with a relative precision of $10^{-3}$. This yields

\begin{equation}
\alpha(r_s)=1+5.888\exp{(-0.07758\; r_s^{0.7923})}.
\end{equation}
The corresponding form of the scattering potential at $r_s = 20$ is also shown in Fig.~1, while Fig.~2 reports a comparison of our results for $g(r)$ with the QMC data up to $r_s = 50$. We have clearly achieved fully quantitative agreement over this whole range of coupling strength. It may be 
remarked that nothing is to be gained by further increasing the relative precision of the fit of the QMC ground-state energy, possibly because of errors due to the finiteness of the QMC sample.

{\it $2D$ fluid of charged fermions}. We carry over Eqs.~(4) and (6) in our calculations of the Bose-like term in Eq.~(1) for the $2D$ electron gas. With regard to the Fermi term $W_F(r)$ in Eq. (1), we 
adopt the same criteria in determining its form as in Ref.~[8]. An important requirement is that Eq.~(1) should give the exact fermion fermion distribution function when one goes to the weak-coupling limit $r_s\rightarrow 0$, when $g(r)$ becomes the Hartree-Fock pair distribution function $g_{HF}(r)$. The Fermi term in the scattering potential is then determined by the Hartree-Fock structure factor $S_{HF}(k)$ according to 

\begin{equation}
W_F(k)=\frac{\hbar^2}{m}{\rm FT}\left[ \frac{\nabla^2\sqrt{g_{HF}(r)}}{\sqrt{g_{HF}(r)}}\right]+\frac{\hbar^2k^2}{4m}\left[\frac{S_{HF}(k)-1}{S_{HF}(k)}\right]^2[2S_{HF}(k)+1]-\alpha(r_s)W_B^{(3)}(k)|_{S(k)=S_{HF}(k)}.
\end{equation}

Here $S_{HF}(k)=\sum_{\sigma}[1+sgn(\sigma)\zeta]~S_{HF}^{\sigma\sigma}(k)/2$ with
\begin{equation}
S^{\sigma\sigma}_{HF}(k)=(2/ \pi)\left[\sin^{-1}(k/2k_{F\sigma})+(k/ 2k_{F\sigma})\sqrt{1-(k/ 2k_{F\sigma})^2}\right]\theta(2k_{F\sigma}-k)+\theta(k- 2k_{F\sigma})
\end{equation}

where $\sigma=\pm1$ denotes the electron spin, $\zeta = |n_{\uparrow}-n_{\downarrow}|/n$ is the spin polarization, $k_{F\sigma}=\sqrt{2}[1+{\rm sgn}(\sigma)\zeta]^{1/2}/(r_s a_B)$ and $\theta$ is the Heaviside step function. The rationale behind Eq.~(8) is as in Ref.~[8]: the first 
term on the RHS ensures that the Hartree-Fock limit is correctly embodied into the theory, and the second and third term ensure that the Bose-like scattering potential is suppressed for parallel-spin electrons at weak coupling~[14].

Our numerical results for the pair distribution function of the $2D$ electron gas in the 
paramagnetic state $(\zeta =0)$ and in the fully spin-polarized state $(\zeta =1)$ are compared with the QMC data of Ref.~[15] in Fig.~3. We have clearly achieved fully quantitative agreement 
with the QMC data up to large values of the coupling strength $r_s$.

We can then confidently calculate the ground-state energy $\varepsilon(r_s,\zeta)$ of the $2D$ electron gas 
as a function of $r_s$ in these two states of magnetization, using an integration over the coupling strength according to the expression

\begin{equation}
\varepsilon(r_s,\zeta)=\frac{(1+\zeta^2)}{r_s^2}+\frac{1}{2}\int_{0}^{1} \frac{d\lambda}{\lambda}~\int \frac{d {\bf k}}{(2 \pi)^2}~v^{(\lambda)}_k~[S_{\lambda}(k)-1]
\end{equation}
(in Rydberg units). The results are compared with the QMC data of Attaccalite {\it et al}.~[7] in Fig.~5. There cleary is a discrepancy with the data in the absolute values of the energy, but this affects 
in essentially the same manner the two phases so that we find a transition from the paramagnetic to the fully spin-polarized fluid at $r_s\approx 24$, in excellent agreement with the QMC data. As to the 
nature of the phase transition, within the accuracy of our calculations it could be either a weakly first-order one or a continuous transition occurring in an extremely narrow range of values of $r_s$. 
This is shown in Fig.~4, where we also report our results for the ground-state energy as a function of $r_s$ at several values of the spin polarization $\zeta$.

{\it Conclusions}. In summary, we have presented a model which quantitatively predicts the two-body 
correlations in both the $2D$ charged-boson fluid and the $2D$ electron gas using as the only input the QMC data for the ground-state energy of the boson fluid as a function of the coupling 
strength $r_s$. The essential physical idea that underlies the model is that differences arising from the statistics are disappearing as the fermionic or bosonic fluid is brought into the strong coupling regime, where the Coulomb repulsions suppress close encounters of pairs of particles~[16]. However, exchange between parallel-spin fermions must be properly accounted for in the weak-to-intermediate coupling regime.

	The model has allowed us to reproduce the quantum phase transition that has been found to occur in the QMC studies of the $2D$ electron gas, essentially starting from the basic Coulomb Hamiltonian. Within the accuracy of our model the transition could be either a weakly discontinuous transition or a continuous one occurring in a very narrow range of coupling 
strength. The latter type of quantum phase transition is found to occur in the $3D$ electron gas~[17].

\vspace{0.5 cm}

\noindent {\bf Ackowledgements}
 We are very grateful to Dr. S. Moroni for sending to us the QMC 
results for $g(r)$ used in Fig.~3. This work was partially funded by MIUR under the PRIN2001 and 
PRIN2003 Initiatives.
\newpage

{\bf References}
\vspace{0.2 cm}

[1] See {\it e.g.} P. Harrison, Quantum Wells, Wires and Dots (Wiley, Chichester, 2002).

[2] See {\it e.g.} N. H. March, Electron Correlation in the Solid State (Imperial College, London,	1999).
 
[3] D.M. Ceperley, B.J. Alder, Phys. Rev. Lett. 45 (1980) 566.

[4]	B. Tanatar, D.M. Ceperley, Phys. Rev. B 39 (1989) 5005.

[5]	F. Rapisarda, G. Senatore, Austr. J. Phys. 49 (1996) 161.

[6]	D. Varsano, S. Moroni, G. Senatore, Europhys. Lett. 53 (2001) 348.

[7]	C. Attaccalite, S. Moroni, P. Gori-Giorgi, G. B. Bachelet, Phys. Rev. Lett. 88 (2002)	256601.

[8]	B. Davoudi, R. Asgari, M. Polini, M.P. Tosi, Phys. Rev. B 68 (2003) 155112.

[9]	E. Krotscheck, M. Saarela, Phys. Rep. 232 (1993) 1.

[10] V. Apaja, J. Halinen, V. Halonen, E. Krotscheck, M. Saarela, Phys. Rev. B         55 (1997)12925.

[11]	S. De Palo, S. Conti, S. Moroni, Phys. Rev. B 69 (2004) 035109.

[12]	A.W. Overhauser, Can. J. Phys. 73 (1995) 683.

[13]	R.A. Smith, A. Kallio, M. Puoskari, P. Toropainen, Nucl. Phys. A328 (1979)        186.

[14]	A. Kallio, J. Piilo, Phys. Rev. Lett. 77 (1996) 4237.

[15]  P. Gori-Giorgi, S. Moroni, G. B. Bachelet, cond-mat/0403050

[16]	M.L. Chiofalo, S. Conti, M.P. Tosi, Mod. Phys. Lett. B 8, (1994) 1207.

[17]	F.H. Zong, C. Liu, D.M. Ceperley, Phys. Rev. E 66 (2002) 036703.

\newpage
\begin{figure}
\begin{center}
\includegraphics[scale=0.8]{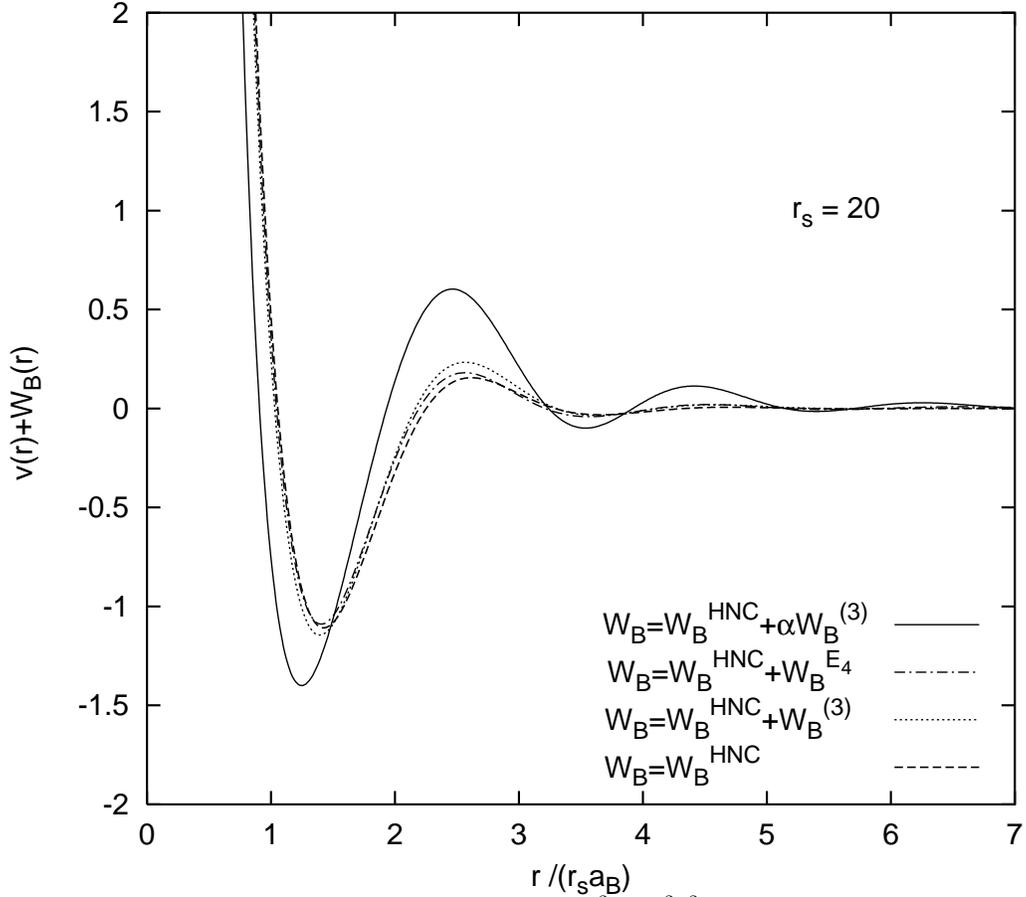}
\caption{The effective Bose-like scattering potential (in units of $\hbar^2/(mr_s^2a_B^2)$) at $r_s = 20$, as a 
function of distance $r$ (in units of $r_s a_B$ ). The four cases refer to the HNC (Eq.~(2)), to the 
inclusion of a contribution of elementary diagrams ($W_B(r)=W_B^{HNC}(r)+W_B^{E_4}(r)$ , Eq.~(3)) and of 
three-body correlations ($W_B(r)=W_B^{HNC}(r)+W_B^{(3)}(r)$ , Eq. (4)), and to the empirically adjusted 
choice in Eq.~(6).
}\label{f1}
\end{center}
\end{figure}

\begin{figure}
\begin{center}
\includegraphics[scale=0.8]{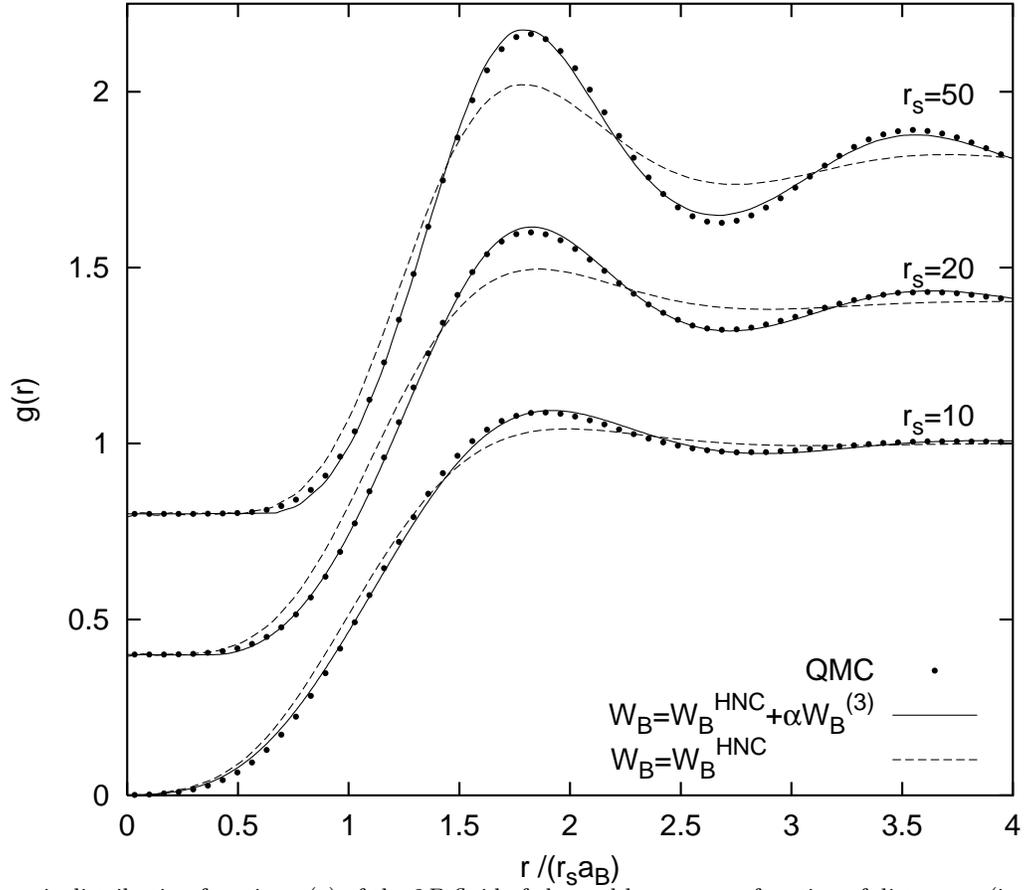}
\caption{The pair distribution function $g(r)$ of the $2D$ fluid of charged bosons as a function of 
distance $r$ (in units of $r_s a_B$) at $r_s$ = 10, 20, and 50. The theoretical results (full lines) are 
compared with QMC data by De Palo {\it et al}.~[11] (dots). The dased lines show the results obtained in the HNC/0 scheme. The results have been shifted upwards by 0.4 and 0.8 for clarity.
}
\label{f2}
\end{center}
\end{figure}

\begin{figure}
\begin{center}
\includegraphics[scale=0.6]{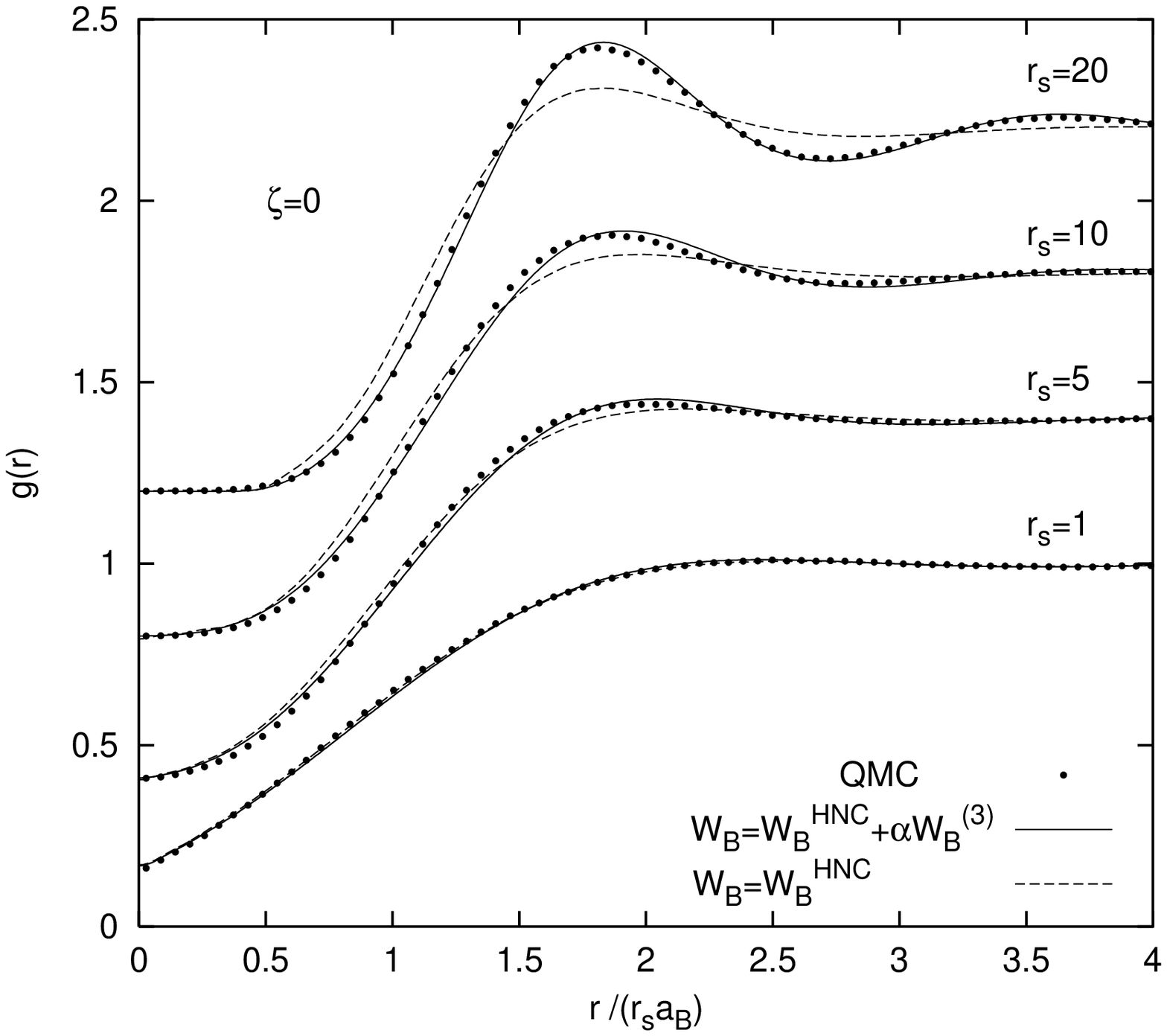}
\includegraphics[scale=0.6]{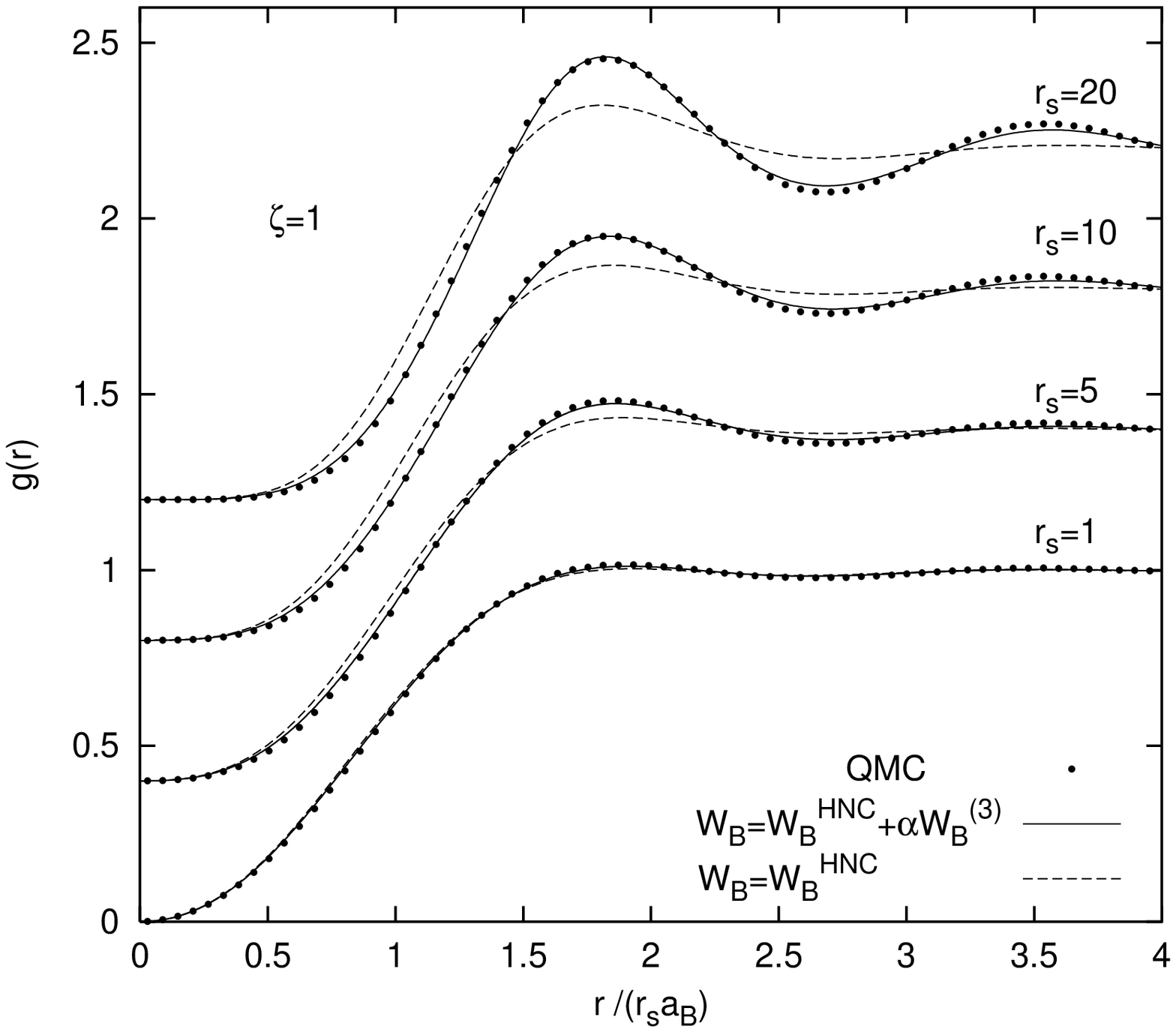}
\caption{The pair distribution function $g(r)$ of the $2D$ electron gas at $r_s =$ 1, 5, 10, and 20 in the paramagnetic state (top panel) and in the fully spin-polarized state (bottom panel), as a function 
of distance $r$ (in units of $r_s a_B$ ). The theoretical results (full lines) are compared with QMC data by Gori-Gorgi {\it et al.}[15] (dots). The dashed lines show the results obtained in the FHNC/0 scheme. The 
results have been shifted upwards by 0.4, 0.8 and 1.2 for clarity.}\label{f3}
\end{center}
\end{figure}

\begin{figure}
\begin{center}
\includegraphics[scale=0.8]{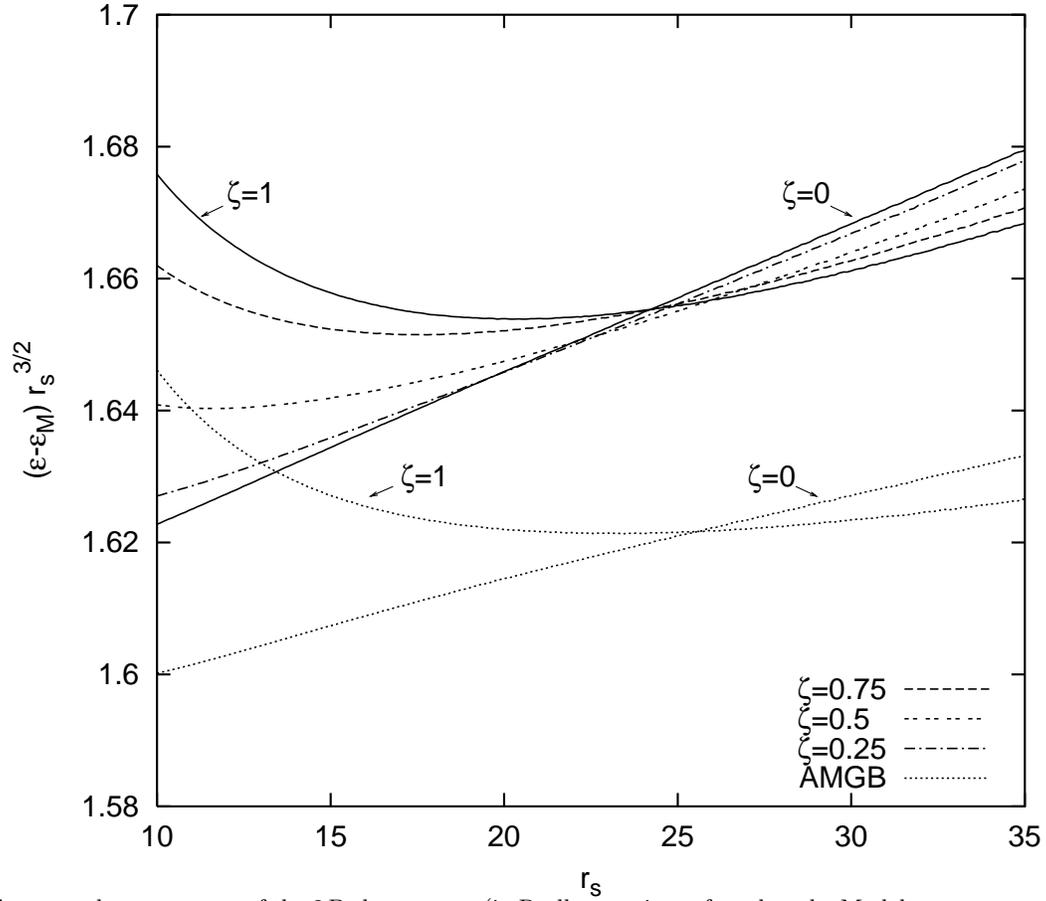}
\caption{The ground-state energy of the $2D$ electron gas (in Rydberg units, referred to the Madelung energy  $\varepsilon_M=-2.2122/r_s$  and multiplied by  $r_s^{3/2}$) as a function of the coupling strength $r_s$ . The full lines show the theoretical results for the paramagnetic state and the fully 
spin-polarized state, while the dots report QMC data from Attaccalite {\it et al}.~[7]. The other three 
curves are theoretical results for states of partial spin polarization, as indicated in the figure.
}
\label{f4}
\end{center}
\end{figure}
\end{document}